\documentstyle[12pt]{article}

\newcommand{\vv}[1]{\mbox{\boldmath{$#1$}}}

\title{Scaling Behavior of Level Statistics in Quantum Hall Regime}

\author{
Yoshiyuki {\sc Ono}, Tomi {\sc Ohtsuki}$^1$ and Bernhard {\sc Kramer}$^2$
}

\begin{document} 

\sloppy
\maketitle

\bigskip
\begin{center}
{\it 
Department of Physics, Toho University, \\
Miyama 2 - 2 - 1, Funabashi, Chiba 274 \\
$^1$Department of Physics, Sophia University, \\
Kioi-cho 7--1, Chiyoda-ku, Tokyo 102 \\
$^2$I. Institut f\"ur Theoretische Physik, Universit\"at Hamburg, \\
D--20355 Hamburg, Germany
}
\end{center}

\medskip

\centerline{(received: February 14, 1996)}

\bigskip
{\small 
The scaling property of level statistics in the quantum Hall regime, i.e. 
2D disordered electron systems subject to strong magnetic fields, is analyzed 
numerically in the light of the random matrix theory. The energy dependences 
of the effective level repulsion parameter, the two level correlation, the 
GUE-GOE crossover parameter, and the rigidity (or $\Delta_3$-statistics) of 
the level distributions are investigated for different system sizes by 
unfolding the original data and by dividing the unfolded spectrum into 
small regions. It is shown that the critical exponent of the localization 
length as a function of energy can be determined through the energy dependence 
of the level statistics.  
The analyses are carried out not only for the lowest Landau band (LB) but 
also for the second lowest LB. Furthermore the effect of finite range of 
disordered potential is studied. The short-ranged potential case in the 
second lowest LB is found to be pathological as in other studies of 
critical behavior, and it is confirmed that this pathological behavior 
is improved in the case of disordered potential with finite ranges. 
}

\newpage

\section{Introduction} 

The study of electronic states in disordered systems is inevitable to 
understand electronic transport properties 
in conducting materials. Most directly the electronic states are described 
by wave functions. In recent researches, however, the importance of energy 
level distributions is pointed out. Since the wave functions and the 
energies are both obtained by solving the eigenvalue problem for 
the disordered Hamiltonian, it is quite natural to expect that the energy 
spectrum involves some informations on the eigenfunctions. 

In disordered systems all the spatial symmetries which may exist in pure 
systems are lost, and only the fundamental symmetries under the operation of 
the time reversal can survive. The importance of such fundamental 
symmetries in the problem of the Anderson localization were first pointed out 
by Wegner\cite{wegner} and Hikami, Larkin and Nagaoka,~\cite{HLN} 
in the treatment of weakly localized regime. 
The relation between the level statistics and the 
fundamental symmetry was originally proposed in nuclear physics in order 
to explain complicated energy spectra in heavy 
nuclei,\cite{wigner,dyson,porter,mehta} and then applied to the problems 
of metallic fine particles\cite{kubo62,goreli} and 
chaos.~\cite{gutzwil,haake}  These level statistics 
are treated within the random matrix theory(RMT),~\cite{porter,mehta} 
where the energy levels 
are mimicked by eigenvalues of random Hamiltonian matrices whose elements 
are chosen randomly. The RMT tells us that the level statistics have 
quite universal properties depending only on the fundamental symmetry of 
the system. If the system is time reversal invariant, there are two 
universality classes; {\it orthogonal} when there is no spin-orbit 
interaction or the spin of the particle is integer and {\it symplectic} when 
the spin of the particle is half odd-integer and there exists a spin-orbit 
interaction. When the time reversal symmetry is violated {\it e.g.} 
by the presence 
of an external magnetic field or magnetic impurities, the corresponding 
universality class is {\it unitary}. 

The usual random matrices and the Hamiltonian matrices describing disordered 
electrons are different in the sense that the matrix elements in the former 
have no correlation while those in the latter have rather strong correlations. 
This will be clear if we remind ourselves of the Anderson model where the 
diagonal elements of the Hamiltonian matrix are uncorrelated random numbers 
but the off-diagonal elements takes 0 or a finite value equal to the 
nearest-neighbor transfer integral in a regular way. Because of this 
correlation, the Anderson model can yield both of localized and delocalized 
states. It is not a trivial task to connect the correlations in matrix 
elements to the localization-delocalization properties of electronic states, 
since the apparent correlations of matrix elements depend on the choice of 
basis functions. It is usually believed that the level statistics of extended 
states are well described by the RMT using Gaussian ensembles.~\cite{efetov} 
The characteristic of the level statistics in the extended regime is the 
strong level repulsion, which can be expressed by the behavior $t^\beta$ of 
the level spacing distribution function $P(t)$ in the small $t$(= spacing) 
region. It is well-known that the level repulsion parameter $\beta$ 
takes 1, 2 and 4 for the Gaussian orthogonal ensemble (GOE), the Gaussian 
unitary ensemble (GUE) and the Gaussian symplectic ensemble (GSE), 
respectively.~\cite{mehta}  This fact represents that the level correlation 
in the extended regime is rather strong. On the other hand the levels in the 
localized regime has no correlation reflecting negligible spatial overlap of 
wave functions of different localized states. The level statistics in this 
case become Poissonian.~\cite{mehta} 

From these behavior of the level statistics it is clear that the 
localization-delocalization properties of disordered systems can be 
analyzed in terms of 
level statistics. First attempt along this line was carried out by Shklovskii 
and co-workers.~\cite{shklo} 
They have analyzed the level spacing distribution in 3D 
Anderson model by changing the system size and the strength of disorder, 
and have found that the level spacing distribution function satisfies 
a certain scaling property. From this scaling property they could derive 
the critical exponent of the localization length at the metal-insulator 
transition. The obtained value of the exponent is consistent with other 
studies such as the real space scaling analysis.~\cite{mack,kramac93}  
A similar scaling behavior 
in the quantum Hall effect regime, i.e. in 2D disordered electron systems 
subject to strong magnetic fields has been found by Ono and 
Ohtsuki,~\cite{oo93,oo94} who 
have found the localization length exponent for the lowest Landau band 
with short-ranged impurity potentials, which is also consistent with 
other studies.~\cite{aoan1,huckek,huob} The level spacing distribution 
at the center of the lowest Landau band was studied by Huckestein and 
Schreiber,~\cite{huckes} who found a distribution quite near to that of GUE, 
though they did not discuss the scaling behavior. The level statistics in 
3D systems without time reversal symmetry have been investigated by Hofstetter 
and Schreiber.~\cite{hofstet3} 
Particularly, they have paid attention to the level statistics at the 
transition, and found that, as far as the level statistics are concerned, 
the behavior at the transition seems not to be affected by the violation of 
the time reversal symmetry, namely that the orthogonal and unitary cases show 
the same level distribution at the transition point, at least in 3D.  The 
conclusion that the two universality classes show the same critical level 
statistics is still controversial, since in the above analysis 
the time reversal symmetry is violated by introducing Aharonov-Bohm type phase 
factor which can affect only the delocalized states and have no effect on the 
localized states.  In order to resolve this controversy, studies on the system 
with real magnetic fields will be necessary. 

Level statistics are studied also analytically.  The results by 
Kravtsov {\it et al.}~\cite{kravtsov} are particularly interesting. 
According to their result, the behavior of the 
level spacing distribution for large spacing region at the transition is 
determined by the localization exponent. Therefore, by analyzing the large 
spacing tail structure of the spacing distribution function at the critical 
point, it is possible to derive the critical exponent of the localization 
length without changing the system size. It is confirmed that this is the case 
at least for 3D orthogonal systems. It will be discussed in this paper that 
the critical level spacing distribution has a size independent form, but that 
a similar analysis as suggested by Kravtsov {\it et al.} does not yield the 
correct exponent.~\cite{oo95} 

The existence of the critical level distribution is understood as follows. 
Let us assume the level distribution is characterized by a certain 
parameter $A$, which takes $A_{\rm loc}$ for localized regime and 
$A_{\rm ext}$ for extended regime. When the system size is infinite, this 
parameter will change abruptly from $A_{\rm loc}$ to $A_{\rm ext}$ as a 
function of energy or disorder strength at the critical point (see 
Fig.~\ref{Fig:1}). 
When the system size is finite, however, the change 
will be gradual.  It is quite natural that the size ($L$) dependence of this 
parameter is scaled by the characteristic length $\xi$ which diverges as a 
function of energy or disorder ($X$) at the transition point ($X_{\rm c}$) 
with the critical exponent $\nu$.  That is 
\begin{equation}
A(L,X) = f(L/\xi(X)) , \hskip 1cm \xi(X) \sim |X-X_{\rm c}|^{-\nu} .
\label{eq1.1}
\end{equation}
At the cricical point where $\xi$ is infinite, $A$ is independent of the 
system size as is seen in Fig.~1. 
This scale invariance at the transition will be applied for any parameter, 
as far as it characterizes the difference of the level statistics 
between the localized and extended regimes. This means that there exits a 
critical scale independent distribution, as long as the above mentioned 
scaling hypothesis is valid. The validity of this hypothesis can be checked by 
analyzing numerical data. 
\begin{figure}
\caption{Typical scaling behavior of a parameter characterizing 
localization-delocalization transition.}
\label{Fig:1}
\end{figure}

In this paper we discuss the scaling behavior of level statistics in the 
quantum Hall regime. Not only the level spacing distribution but also the 
$\Delta_3$-statistics, two level correlation function and so forth are found 
to obey the same scaling behavior. We also study the effects of potential 
range on the level statistics. In previous works, mainly the lowest Landau 
band was studied, but in the present work we discuss also the level statistics 
in the second lowest Landau band.~\cite{ono94} 

\section{Model and Formulation} 

The system discussed in this paper is a two-dimensional non-interacting 
electron gas in a strong magnetic field and a random potential. The magnetic 
field is assumed to be so strong that the Landau band separation is much 
larger than the Landau band width due to scattering by the random potential. 
As for the random potential $V(\vv{r})$ we assume that of the Gaussian type 
with the following properties, 
\begin{eqnarray}
<V(\vv{r})> & = & 0 , \label{eq2.1} \\
<V(\vv{r}+\vv{a})V(\vv{r})> & = & {{u^2} \over {2\pi s^2}}
\exp (-\vv{a}^2/2s^2) , \label{eq2.2}
\end{eqnarray}
where $s$ and $u$ represent the correlation length and the strength 
of the potential, respectively. They are related to the lowest Landau band 
width $\Gamma_{\rm SCBA}^{(0)}$ calculated within the self-consistent 
Born approximation (SCBA), as follows,~\cite{ando1}
\begin{equation}
\Gamma_{\rm SCBA}^{(0)} = \sqrt{{2u^2} \over {\pi(l^2 + s^2)}}, 
\end{equation}
where $l$ is the magnetic length. The SCBA band width is independent of 
Landau band index in the case of short-ranged potentials ($s=0$, i.e. 
$\delta$-correlated), but depends on it in the case of finite ranges 
($s \neq 0$).~\cite{ando1} 
In the present treatment, however, such 
detailed dependences of the band width on various parameters are not 
important, since the characteristics of the unfolded levels are considered 
to be independent of the absolute value of the band width (see below). 

In order to obtain the energy spectra, one must first calculate 
the Hamiltonian matrix elements by introducing appropriate basis functions. 
In the present situation a set of Landau functions is the most appropriate 
basis. As discussed in refs.~\cite{huckek} and \cite{oksok}, 
it is possible to generate random 
elements of the Hamiltonian matrix which reflect the essential features 
of the Landau functions as basis and of the random potential as described 
in eqs. (\ref{eq2.1}) and (\ref{eq2.2}).  Details of preparing the matrix 
elements have been described in refs.~\cite{huckek} and \cite{oksok}, 
and therefore we do not repeat them here. As in ref.~\cite{oksok}, 
the periodic boundary conditions are assumed for both 
directions in order to avoid edge effects. In the present work, we treat 
not only the lowest Landau band but also 
the second lowest one to study the effect of different basis 
sets.~\cite{liusarma,onofuku} 
In fact the case of the short-ranged potential and the second lowest Landau 
band is pointed out to behave pathologically in the study of the localization 
length by the real space scaling analyses.~\cite{aoan1,liusarma,mieck,hucke94}

The density of states for each Landau band takes bell shape in most cases. 
This means that the average level spacing changes as a function of energy. 
As is well-known, a procedure called ^^ ^^ unfolding'' maps the original 
energy levels $\{ E_i\} $ to a new set of levels $\{ x_i\} $ which has a 
constant average spacing independent of the level region.~\cite{mehta} 
If we represent the integrated density of states by $N(E)$, then the 
unfonlding is carried out by 
\begin{equation}
x_i = N(E_i) . 
\end{equation}
It is straightforward to show that the density of states and therefore the 
average spacing for the new levels are constant.~\cite{oksok}  
Throughout the paper, we discuss the statistics of the unfolded levels. 

The Hamiltonian matrix size increases when we consider larger systems. 
If we denote the matrix size by $N_{\rm mat}$ and the system size by 
$L\times L$ (i.e. the linear dimension, $L$), they are related to each 
other in the following form, 
\begin{equation}
L = \sqrt{2\pi N_{\rm mat}}l . \label{eq2.5}
\end{equation}
The matrix sizes treated here are 
$N_{\rm mat} = 200, 400, 600$ and 800. The number of samples for each size 
is chosen to be $N_{\rm sample} = 288,000/N_{\rm mat}$ so that the total 
number of energy levels may be 288,000. 
The potential range discussed here covers 0 to 2$l$.


\section{Level Spacing Distribution}

As is clear from the definition of the unfolding procedure, the unfolded 
levels are distributed between 0 and $N_{\rm mat}$, and the average spacing is 
equal to unity.~\cite{oksok} It is well known 
that the truly delocalized states 
are possible only at the Landau band center. We shift the unfolded levels 
by $N_{\rm mat}/2$ in order to make $x=0$ correspond to the band center. 
As mentioned in \S~1, the statistics of levels of delocalized 
states and those of localized ones are qualitatively different. In order 
to see the transition between localized and delocalized characteristics, 
we consider the energy dependent level statistics.  In practice we take 
a region of $x$ with a width of $0.1N_{\rm mat}$, and discuss 
the level statistics 
within that region. The region is represented by its central value. 
By changing this central value, we can discuss the level statistics dependent 
on the energy region. In the case of the quantum Hall regime, the Landau 
band center is always the critical point in the sense of the Anderson 
localization. It is, therefore, not possible to see the transition by 
changing the strength of disorder as in the case of the 3D orthogonal 
case.~\cite{shklo,mack,kramac93} This is the reason why we discuss the energy dependence. 
\begin{figure}
\caption{Examples of level spacing distributions for the lowest 
Landau band with $\delta$-correlated potential; (a) $x=0$ and (b) $x=-240$, 
$N_{\rm mat}$ being 600 in both figures.}
\label{Fig:2}
\end{figure}

The examples of level spacing distributions $P(t)$ ($t$ the spacing of the 
unfolded levels) are given in Fig.~\ref{Fig:2} for the case 
of the lowest Landau band with $\delta$-correlated potential. 


\subsection{Effective level repulsion parameter}

As discussed in ref.~\cite{oksok}, 
one of the parameters characterizing these distributions is the effective 
level repulsion parameter $\beta$, which is thought to be 2 for the extended 
regime in the unitary symmetry and 0 for the localized regime. 
Several interpolation formulae are proposed,~\cite{brody,izrailev} 
which describe intermediate distributions between those for the Gaussian ensembles, 
\begin{eqnarray}
P_{\rm GOE} (t) & = & {{\pi} \over 2}t\exp(-{{\pi} \over 4}t^2), \nonumber \\
 & & \mbox{\rm for Gaussian orthogonal ensemble}, \\ 
P_{\rm GUE} (t) & = & {{32} \over {\pi^2}}t^2 \exp(-{4 \over {\pi}} t^2), 
\nonumber \\
& & \mbox{\rm for Gaussian unitary ensemble}, \\ 
P_{\rm GSE} (t) & = &   {{2^{18}} \over {3^6 \pi^3}}t^4 \exp(-{{64} \over 
{9\pi}}t^2), \nonumber \\ 
& & \mbox{\rm for Gaussian symplectic ensemble},  
\end{eqnarray}
and the Poissonian, 
\begin{equation}
P(t) = {\rm e}^{-t} 
\end{equation}

We apply the simplest interpolation formula proposed by Brody~\cite{brody} 
and extended to cover the case $\beta > 1$.~\cite{oksok} 
\begin{eqnarray}
P_{\beta}(t) & = & (1+\beta)a(\beta)t^{\beta} \exp (-a(\beta)t^{1+\beta}), 
\nonumber \\
& & \hskip 25mm {\rm for} \  0 \le \beta \le 1, \label{eq3.5} \\
P_{\beta}(t) & = & c(\beta)t^{\beta} \exp (-b(\beta)^2t^2) , 
\hskip 5mm {\rm for} \ \beta > 1, \label{eq3.6}
\end{eqnarray}
where 
\begin{eqnarray}
a(\beta) & = & \left[ \Gamma \left({{2+\beta} \over {1+\beta}}\right) 
\right]^{1+\beta} , \\ 
b(\beta) & = & {{\beta \Gamma({{\beta} \over 2})} \over 
{2\Gamma({{1+\beta} \over 2})}} , \\ 
c(\beta) & = & {{2b(\beta)^{1+\beta}} \over {\Gamma ({{1+\beta} \over 2})}}. 
\end{eqnarray}
Even if we use other more complicated formulae such as that proposed by 
Izrailev,~\cite{izrailev} the final result is found not to be affected. 
Although there is some argument that the small $t$ behavior of 
$P(t)$ for delocalized states is described always by $t$, $t^2$, or $t^4$ 
depending on the basic symmetry of the system, we consider the effective 
level repulsion parameter $\beta$ obtained by fitting the data to an 
interpolation formula like eqs. (\ref{eq3.5}) and (\ref{eq3.6}) can describe 
to some extent the deviation of the numerical data from the distribution for 
Gaussian ensemble, even if, precisely speaking, the fitting might be poor.  
In fact, Shklovskii {\it et al.}~\cite{shklo} 
chose an integral of $P(t)$ in a certain fixed region by noting that the 
intermediate distributions cross at a point near $t=2$ in the 3D orthogonal 
case.  Unfortunately there is no such common crossing point in the case of the 
quantum Hall regime. 
\begin{figure}
\caption{The scaling behavior of the effective level repulsion parameter 
in the case of the lowest Landau band with two different correlation lengths 
of the random potential; (a) $s = 0$ and (b) $s = 2l$ ($l$: the magnetic 
length).}
\label{Fig:3}
\end{figure}

Two examples of the scaling behavior of the effective level repulsion 
parameter, obtained as mentioned above, is shown in Fig.~\ref{Fig:3} for the 
case of the lowest Landau band with different correlation lengths of random 
potential. From this behavior we can deduce 
the critical exponent of the localization length $\nu$: Energy dependence 
of the localization length $\xi$ is assumed to be described by
\begin{equation}
\xi \propto |E-E_N|^{-\nu}, 
\end{equation}
where $E_N$ represents the center of the $N$-th Landau band. 
Then $\beta$ as the function of $x$ and $N_{\rm mat}$ is 
\begin{eqnarray}
\beta(x, N_{\rm mat}) & = & f(|x/N_{\rm mat}|^{-\nu}N_{\rm mat}^{-1/2}) 
\nonumber \\
& = & \tilde{f}(x/N_{\rm mat}^{1-1/2\nu}) ,  \label{eq3.10a}
\end{eqnarray}
from eqs.~(\ref{eq1.1}) and (\ref{eq2.5}). 
The obtained critical exponent $\nu = 2.4$ for the case of the lowest 
Landau band with $\delta$-correlated potential is consistent with that of 
different previous studies.~\cite{aoan1,huckek,huob,liusarma,kramack} 

Similar scaling behaviors were investigated for different potential 
ranges and different Landau bands. 
Particularly it is confirmed that the case of the second lowest Landau band 
with the $\delta$-correlated potential is pathological (see Fig.~\ref{Fig:4}) 
as already found in previous studies of localization length 
exponents.~\cite{aoan1,liusarma,mieck,hucke94} Furthermore, at least 
within the present analysis, the exponent depends on the potential range 
and universality was not confirmed, while the universality within the 
lowest Landau band has been confirmed in the real space scaling 
analysis.~\cite{hucke92} The critical exponents obtained in this analysis 
are summarized in Table~\ref{Tab:I}. It is found in general that the 
increase of the potential correlation length leads to smaller exponent. 
It is also found that the pathological behavior of the second lowest Landau 
band with short-ranged potential can be improved by introducing finite 
correlation length.~\cite{hucke92} 
\begin{figure}
\caption{Scaling behavior of the effective level repulsion parameter in 
the second lowest Landau band with the correlation length of the potential 
(a) $s=0$ and (b) $s=l$ ($l$ the magnetic length). }
\label{Fig:4}
\end{figure}

\begin{table}
\caption{Critical exponents obtained from the scaling behavior 
of the effective level repulsion parameter $\beta$.}
\label{Tab:I}
\begin{tabular}{c|ccc|cccccc} \hline
Landau band index &   & 0   &      &   &        &     1  &  &  &  \\ \hline
potential range $s$    & 0 & $l$ & 2$l$ & 0 & 0.5$l$ & 0.7$l$ & 0.8$l$ & $l$ & 
2$l$ \\ \hline
exponent $\nu$ & 2.4 & 2.0 & 1.5 & 7.1 & 5.0 & 2.5 & 2.4 & 1.7 & 1.3 \\ 
\hline
\end{tabular}
\end{table}

The lack of universality within the present analysis may be partly due to 
the finite width of the energy window. In order to check the effect of the 
finiteness of the window, we might have to take many more samples and make the 
window width much smaller. We have made the window width half without 
increasing the sample number, but we have got no change in the scaling 
behavior of the effective level repulsion parameter, at least to this 
stage.


\subsection{GUE--GOE crossover parameter}

In ref.~\cite{oksok} 
it has been discussed that the level spacing distribution in the QH regime 
can be characterized by the GUE-GOE crossover parameter. 
According to the semiclassical argument by Argaman, Imry and 
Smilansky,~\cite{argaman} 
a smaller level separation corresponds to a longer period of the semiclassical 
motion and therefore to a larger orbit. The larger the orbit, the stronger the 
effect of the magnetic field will be. From these consideration it is expected 
that the small $t$ behavior of $P(t)$ resembles that of GUE while the large 
$t$ behavior is similar to that of GOE. The crossover from GUE in small $t$ 
region to GOE in large $t$ region can be described by a model proposed by 
Pandey and Mehta~\cite{pandey} and utilized by Dupuis and 
Montambaux~\cite{dupuis} to explain the level 
spacing distribution in a small metallic ring pierced by a magnetic flux. 

In Pandey and Mehta's model,~\cite{pandey} the random Hamiltonian matrix 
is expressed in 
a linear combination of symmetric and anti-symmetric matrices with a 
coefficient i$\alpha$ ($\alpha$; real) for the latter. The case with 
$\alpha=0$ corresponds to GOE and the case with $\alpha=1$ to GUE. The 
level spacing distribution function in the intermediate case is 
obtained by considering 2$\times$2 matrices. The resulting distribution 
function is given by~\cite{french} 
\begin{equation}
P_{\alpha}(t) = {t \over {4v^2\sqrt{1-\alpha^2}}}\exp(-t^2/8v^2){\rm Erf}(wt) 
\label{alphafit} 
\end{equation}
with 
\begin{eqnarray}
v & = & \sqrt{\pi \over 8} \left( \alpha +
{1 \over {\sqrt{1-\alpha^2}}}{\rm arctan}
{{\sqrt{1-\alpha^2}} \over \alpha} \right)^{-1} , 
\\ 
w & = & \sqrt{{1-\alpha^2} \over {8\alpha^2v^2}} . 
\end{eqnarray}

In fact, as far as the effective level repulsion parameter $\beta$ is larger 
than unity, the level spacing distribution data can be fitted to this interpolation formula and the crossover parameter is estimated as a function 
of energy. The energy and size dependence of $\alpha$ satisfies 
the scaling relation as is seen in Fig.~\ref{Fig:5}. Note that data near the 
band edges cannot be used in this analysis. 
\begin{figure}
\caption{Scaling behavior of $\alpha$ in the lowest Landau band with 
$\delta$-correlated potential.}
\label{Fig:5}
\end{figure}

It is worthwhile to note that this scaling relation is explained by the same 
critical exponent which is found for the effective level repulsion parameter. 
Unfortunately, however, the crossover fitting is not possible when the maximum 
value of the effective level repulsion parameter $\beta$ is less than unity, 
which occurs for long-ranged potentials.


\subsection{Critical distribution}

The behavior of the scale invariant critical distribution function of the 
level spacing at the center of the lowest Landau band with 
$\delta$-correlated potential has been discussed in ref.~\cite{oo95} 
Therefore we mention here that the critical distribution is well fitted 
over all the region of $t$ by the expression,
\begin{equation}
P_{\rm crit}(t) = At^2\exp(-Bt^{2-\gamma}), \label{cridis} 
\end{equation}
with $\gamma$ the fitting parameter. The coefficients $A$ and $B$ are 
determined by two normalization conditions, i.e. $<1> = <t> = 1$. 
According to the 
analytic study,~\cite{kravtsov} $\gamma$ is related to the dimension $d$ of 
the system and the critical exponent $\nu$ as
\begin{equation}
\gamma = 1-{1 \over {\nu d}} . \label{eq3.14}
\end{equation}
The best fit value of $\gamma$ in the case of the lowest Landau band 
with short-ranged potential is about 0.35, which yields the critical exponent 
smaller than unity in disagreement with other studies. 

We have studied the critical distributions for different values of the 
potential range and also for the second lowest Landau bands. It was 
confirmed that the critical distributions are scale invariant and can be 
fitted to eq. (\ref{cridis}) quite well as long as the potential is 
short-ranged.  For longer-ranged potentials the single parameter fitting to 
eq. (\ref{cridis}) is not successful, and we have to use a two parameter 
fitting to 
\begin{equation}
P_{\rm crit}'(t) = A't^{\beta}\exp(-B't^{2-\gamma}). 
\label{cridis2} 
\end{equation}
Here again the parameters $A'$ and $B'$ are related to $\beta$ and $\gamma$ 
due to the two normalization condition. In order to show how the potential 
range affects the critical distribution, we give the data of the critical 
level spacing distribution for the case of the second lowest Landau band 
with the potential correlation length $s = 2l$ in Fig.~\ref{Fig:6} along with 
the fitting curve to eq.~(\ref{cridis2}). The values of $\beta$ and $\gamma$ 
obtained by the best fit to eq. (\ref{cridis2}) are summarized 
in Table~\ref{Tab:II}. 
\begin{figure}
\caption{Critical level spacing distribution at the center of the second 
lowest Landau band with the potential correlation length $s=2l$. The solid 
curve is the best fit to eq.~(3.17) with $\beta = 0.66$ and 
$\gamma = -0.58$.}
\label{Fig:6}
\end{figure}

\begin{table}
\caption{Values of $\beta$ and $\gamma$ obtained by fittig 
the critical level spacing distribution at the band center 
to eq.~(3.17) for different cases.}
\label{Tab:II}
\begin{tabular}{c|ccc|cccccc} \hline
Landau band index &   &  0  &      &   &        &    1   &  &  &  \\ \hline
potential range $s$    & 0 & $l$ & 2$l$ & 0 & 0.5$l$ & 0.7$l$ & 0.8$l$ & $l$ & 
2$l$ \\ \hline
$\beta$  & 2.19 & 1.95 & 1.03 & 2.04 & 2.01 & 2.02 & 2.03 & 2.01 & 0.66 \\ 
$\gamma$ & 0.47 & 0.48 & 0.02 & 0.17 & 0.16 & 0.25 & 0.31 & 0.43 & $-$0.58 \\ 
\hline
\end{tabular}
\end{table}

It will be worthwhile to point out that the fitting to eq. (\ref{cridis2}) 
and that to eq. (\ref{alphafit}) are almost indistinguishable as far as 
the latter fitting is possible. 


\section{$\Delta_3$-Statistics}

In the random matrix theory, the rigidity of the level distribution is studied 
in terms of the $\Delta_3$-function.~\cite{mehta} It is defined by
\begin{eqnarray}
\Delta_3 (K, x_0) & = & \min_{A,B} {1 \over K} \int_{x_0-K/2}^{x_0+K/2} 
{\rm d}x\ [M(x) - A x -B]^2  \nonumber \\
& = & <M^2> - <M>^2 -{{12} \over {K^2}}<(x-x_0)M>^2 , \nonumber \\
& &       \label{del3} 
\end{eqnarray}
where $M(x)$ is the integrated density of states for the unfolded levels 
\begin{equation}
M(x) = \int_0^x W(x){\rm d}x \hskip 1cm (W(x); \mbox{\rm the density of 
states}) , \label{M_x} 
\end{equation}
and $<\cdots>$ means
\begin{equation}
<\cdots> = {1 \over K} \int_{x_0-K/2}^{x_0+K/2}{\rm d}x \cdots . 
\end{equation}

If  $\Delta_3$ is a constant independent of $K$, the levels must be 
equidistant. If the levels are uncorrelated and obey the Poisson distribution, 
it can be shown that $\Delta_3$ is proportional to $K$. The exact $K$ 
dependences of $\Delta_3$ for GOE, GUE, GSE and the Poissonian are 
known.~\cite{mehta}

The sample averages are taken for fixed $K$ and $x_0$.~\cite{oksok} 
An example of $\Delta_3(K, x_0)$ calculated from numerical data is shown in 
Fig.~\ref{Fig:7} for the case of the lowest Landau band with short-ranged 
disordered potential.  It is confirmed that at the band centers 
the $K$-dependence of $\Delta_3$ is size independent. This 
means that $\Delta_3$ shows a kind of critical behavior at the band centers. 
Note that it is difficult to fix the energy region in this type of analysis 
since the finite value of $K$ is necessary by 
definition.~\cite{foot2} 
\begin{figure}
\caption{Examples of the $\Delta_3$-statistics in the lowest Landau band 
with $\delta$-correlated potential. Two values of $x_0$ are chosen; $x_0 = 0$ 
(the band center) and $x_0 = 0.2N_{\rm mat}$.  The solid line is the analytic 
form for the GUE.}
\label{Fig:7}
\end{figure}

In order to parametrize the $\Delta_3$-function, we introduce its integral 
in the form, 
\begin{equation}
D_3(x_0) = \int_0^{K_0} {\rm d}K \Delta_3(K, x_0), \label{D_3} 
\end{equation}
choosing $K_0$ as a sufficiently large constant. Then we consider 
the energy ($x_0$) and the size ($N_{\rm mat}$) dependences of 
this integral.  It is easily confirmed that this energy and size dependences 
follow the same scaling relation as other parameters discussed above 
(see Fig.~\ref{Fig:8}).
\begin{figure}
\caption{An example of the scaling behavior of the parameter $D_3$; 
the lowest Landau band with $\delta$-correlated potential. 
For the upper limit of the integral $K_0$, two values are chosen.}
\label{Fig:8}
\end{figure}

The choice of $K_0$ is found not to be crucial for the final scaling 
argument (see Fig.~\ref{Fig:8}). It is confirmed that different choices of 
$K_0$ gives 
the same scaling exponent though the absolute value of $D_3$ depends on $K_0$. 


\section{Two Level Correlation Function}

It has been pointed out that the two level correlation function plays an 
important role to determine the behaviors of physical quantity such as 
response functions.~\cite{altshklo}

By using the density of unfolded levels before sample average $W(x)$, 
the two level correlation function $Y$ is defined as
\begin{equation}
Y(\omega, x_0, N_{\rm mat}) = \overline{W(x)W(x+\omega)}, 
\end{equation}
where the overline means the averages over $x$ and over samples under the 
condition that both levels $x$ and $x+\omega$ are within the region 
$(x_0-N_{\rm mat}/20,x_0+N_{\rm mat}/20)$.~\cite{foot1}
Examples of the two level correlation function in the cases of the two 
different Landau bands with short-ranged disordered potential and with the 
matrix size $N_{\rm mat}=400$ are shown in Fig.~\ref{Fig:9} for $x_0 = 0$ 
and $-$180. 
For large level separation $\omega$ it tends to unity irrespectively of the 
extent of the localization. This is because the densities of states at 
sufficiently separated energies are uncorrelated and the average value of 
the density of unfolded levels is unity by definition.~\cite{oksok} 
On the other hand, when the separation $\omega$ is small, the two 
level correlation function behaves in much different ways depending 
whether the corresponding states are delocalized or strongly localized. 
The exact functional forms of $Y(\omega)$ are known for GOE, GUE, and GSE 
cases,~\cite{mehta} which are believed to describe the level statistics in 
the metallic regime. For example, that for GUE is given by
\begin{equation}
Y_{\rm GUE}(\omega)=1- \left( {{\sin \pi \omega} \over {\pi \omega}}\right)^2 
\end{equation}
In Fig.~\ref{Fig:9} we show $Y_{\rm GUE}(\omega)$ by a thin dashed curve. 
\begin{figure}
\caption{Examples of two level correlation function for the case with 
$N_{\rm mat}=400$ and $x_0=0$ and $-$180; (a) the lowest Landau band 
and (b) the second lowest Landau band. In both cases, the disordered 
potential is $\delta$-correlated.  The dashed line represents 
$Y_{\rm GUE}(\omega)$.}
\label{Fig:9}
\end{figure}

In order to parametrize the energy and size dependence of the two level 
correlation function, we introduce the following integral, 
\begin{equation}
z(x_0,N_{\rm mat})= \int_0^{2} [1-Y(\omega, x_0, N_{\rm mat})] {\rm d}\omega . 
\end{equation}
Here the upper bound of the integral has no special meaning; any value 
can be chosen as far as it is in the region where $Y(\omega)$ is almost 
constant for any value of $x_0$. It is not difficult to confirm that 
this parameter $z(x_0,N)$ obeys the same scaling relation as discussed 
for other parameters characterizing the level statistics in previous 
sections (see Fig.~\ref{Fig:10}). 
\begin{figure}
\caption{Scaling behavior of the parameter $z(x_0,N)$ obtained from 
the two level correlation function $Y(\omega, x_0, N)$ in the lowest 
and the second lowest Landau bands with short-ranged random potential.}
\label{Fig:10}
\end{figure}


\section{Summary and Discussion}

In this paper we have discussed the scaling behavior of the energy and 
system size dependences of the level statistics in 2D disordered electron 
systems subject to strong magnetic fields (the quantum Hall regime) from 
various points of view in the light of the RMT. Not only the level 
statistics in the lowest Landau band but also those in the second lowest 
Landau band are considered. Furthermore the correlation length of the 
disordered potential has been set to various values. 

In treating the level spacing distribution, we have discussed the effective 
level repulsion parameter and the GUE-GOE crossover parameter. Both 
parameters are found to satisfy the same scaling relation. From this scaling 
behavior the critical exponent of the localization length can be estimated. In 
the case of the lowest Landau band with short-ranged potential, the exponent 
is consistent with that obtained by other studies.~\cite{aoan1,huckek,huob} 
It is also confirmed that the second lowest Landau band with short-ranged 
potential is pathological.~\cite{liusarma,mieck,hucke94} The longer 
correlation length of the disordered potential leads to the smaller exponent 
within the present treatment. Namely the universality was not confirmed even 
within the lowest Landau band in contrast to the conclusion of the real space 
scaling analysis.~\cite{hucke92} 

Experimental situation about the universality is 
a bit delicate.  In AlGaAs/GaAs systems, Koch {\it et al.}~\cite{koch} 
obtained the critical exponent $\nu = 2.3$ for three lowest Landau bands and 
claimed to have confirmed the universality. On the other hand, in the case 
of Si-MOS systems, magnetic field and sample dependent exponents are 
reported.~\cite{dolgopolov,shashkin,shinohara} Although one of the 
reasons of the lack of universality, if any, may be the mutual interaction 
of electrons,~\cite{shinohara} the possibility of the potential range 
dependence of the exponent should be reconsidered. In experiments, the energy 
of the carrier cannot be precisely fixed, and therefore the observed 
quantities may be coarse-grained. Even if the rigorous exponent might have a 
universality, the exponent estimated from coarse-grained measurements may not 
apparently satisfy the universality. The present analysis of the critical 
behavior in terms of level statistics will correspond to the measurements 
coarse-grained in the energy region.

The pathological behavior of the second lowest 
Landau band is improved by increasing the correlation length of the potential 
near to magnetic length. If we increase the correlation length up to two 
times the magnetic length, the maximum value of the effective level repulsion 
parameter gets very small. This would not be inconsistent with the notion 
that in the limit of slowly varying potential (so-called percolation limit) 
the electronic states correspond to semiclassical orbits along equipotential 
lines. In this limit the level repulsion will become very small. 

The same scaling behaviors have been found for the $\Delta_3$-statistics and 
the two level correlation function. The fact that various parameters 
characterizing the level statistics satisfy the same scaling relation 
means the existence of the characteristic length determining the size 
dependence of the level statistics. In the present problem it is most 
natural to consider that this length is the localization length. 

The pathological behavior of the second lowest Landau band might be 
understood by introducing a very long irrelevant length as suggested by 
Huckestein.~\cite{hucke94} However the physical meaning of the irrelevant 
field is not clear. Liu and Das Sarma~\cite{liusarma} suggested that the 
spatial symmetry of the basis function of each Landau band might play some 
role in this pathological behavior.  In order to see whether this is the 
case, we should investigate higher Landau bands. 

One of the most important results in the present study is the existence of 
the scale invariant critical statistics.  For the critical level spacing 
distribution, the form eq.~(\ref{cridis}) with eq.~(\ref{eq3.14}) is suggested 
in the analytic study by Kravtsov {\it et al}.~\cite{kravtsov} 
Our numerical data in the case of short-ranged potentials can be fitted to 
it, but cannot give a correct exponent. Furthermore we have to use 
two-parameters fitting in the case of long-ranged potentials. In the case of 
the symplectic symmetry, a similar problem has been pointed out by several 
authors.~\cite{oo95,schwzha,evange95}  In 3D orthogonal case, the critical 
level spacing distribution corresponding to eq.~(\ref{cridis}) could be 
successfully applied~\cite{hofstet2} to some extent. Even in this case, 
however, the existence of the deviation from that expression in the 
large separation region has been pointed out.~\cite{zharek} 
More elaborated analytical studies will be necessary in order to understand 
truly the critical level statistics at the Anderson transition. Present 
numerical work on the quantum Hall regime encourages certainly such studies.

\section*{Acknowledgments} 

Present authors are grateful to Keith Slevin, Isa Zharekeshev and 
T. Kawarabayashi for fruitful discussions. This work was partly financed 
by a Grant-in-Aid for Scientific Research from the Ministry of Education, 
Science and Culture, 07640520. 


\def\pr{Phys. Rev. } 
\def\prb{Phys. Rev. B } 
\def\prl{Phys. Rev. Lett. } 
\def\jpsj{J. Phys. Soc. Jpn. }  
\def\ssc{Solid State Commun. }


\begin{thebibliography}{99} 
%
\bibitem{wegner} F.J. Wegner: Phys. Rev. B{\bf 19} (1979) 783, 
Z. Phys. B{\bf 35} (1979) 207.
%
\bibitem{HLN} S. Hikami, A.I. Larkin and Y. Nagaoka: Prog. Theor. Phys. 
{\bf 63} (1980) 707.
%
\bibitem{wigner} E.P. Wigner: Ann. Math. {\bf 53} (1951) 36, {\bf 62} (1955)
548, {\bf 65} (1957) 203, {\bf 67} (1958) 325. 
%
\bibitem{dyson} F.J. Dyson: J. Math. Phys. {\bf 3} (1962) 140, 157, 166.
%
\bibitem{porter} C.E. Porter (ed.): {\it Statistical theories of spectral
fluctuations} (Academic Press, 1965).  
%
\bibitem{mehta} M.L. Mehta: {\it Random Matrices} 2nd ed. (Academic Press,  
1991). 
%
\bibitem{kubo62} R. Kubo: J. Phys. Soc. Jpn. {\bf 17} (1962) 975. 
%
\bibitem{goreli} L.P. Gor'kov and G.M. Eliashberg: Zh. Exsp. Teor. Fiz. 
{\bf 43} (1965) 1407.
%
\bibitem{gutzwil} M. C. Gutzwiller: {\it Chaos in Classical and Quantum 
Mechanics} (Springer, New York 1990)
%
\bibitem{haake} F. Haake: {\it Quantum Signatures of Chaos} (Springer, Berlin 
1991)
%
\bibitem{efetov} K.B. Efetov: Adv. Phys. {\bf 32} (1983) 53.
%
\bibitem{shklo} B.I. Shklovskii, B. Shapiro, B.R. Sears, P. Lambrianides
and H.B. Shore:  Phys. Rev. {\bf B 47} (1993) 11487.  
%
\bibitem{mack} A. MacKinnon  and B. Kramer: \prl {\bf 47} (1981) 1546,  
Z. Phys. {\bf B 53} (1983) 1.
%
\bibitem{kramac93} B. Kramer and A. MacKinnon: 
Rep. Prog. Phys. {\bf 56} (1993) 1469. 
%
\bibitem{oo93} Y. Ono and T. Ohtsuki: \jpsj {\bf 62} (1993) 3813. 
%
\bibitem{oo94} Y. Ono and T. Ohtsuki: \jpsj {\bf 63} (1994) Suppl. A 158. 
%
\bibitem{aoan1} H. Aoki and T. Ando: \prl {\bf 54} (1985) 831, T. Ando  
and H. Aoki: \jpsj {\bf 54} (1985) 2238.
%
\bibitem{huckek} B. Huckestein and B. Kramer: \ssc {\bf 71} (1989) 445, \prl
{\bf 64} (1990) 1437.
%
\bibitem{huob} Y. Huo and R.N. Bhatt: \prl {\bf 68} (1992) 1375. 
%
\bibitem{huckes} B. Huckestein and L. Schweitzer: in {\it High Magnetic 
Fields in Semiconductor Physics} III ({\it Proc. Int. Conf. Appl. High Mag.  
Fields in Semicond. Phys. 1990 August, W\"urzburg F.R.G.}),
ed. G. Landwehr (Springer, 1992) p. 84. 
%
\bibitem{hofstet3} E. Hofstetter and M. Schreiber: \prl {\bf 73} (1994) 3137.
%
\bibitem{kravtsov} V.E. Kravtsov, I.V. Lerner, B.L. Al'tshuler, and 
A.G. Aronov: \prl {\bf 72} (1994) 888; A.G. Aronov, V.E. Kravtsov and 
I.V. Lerner: JETP Lett. {\bf 59} (1994) 39.
\bibitem{oo95} T. Ohtsuki and Y. Ono: \jpsj {\bf 64} (1995) 4088.
%
\bibitem{ono94} A part of the present work was presented at the International 
Conference on {\it ^^ ^^ High Magnetic Fields in Semiconductor Physics''} 
held at MIT, Boston USA, 1994 August, Y. Ono: in {\it ^^ ^^ High Magnetic 
Fields in the Physics of Semiconductors''}, ed. D. Heiman (World Scientific, 
Singapore, 1995), p. 260.
%
\bibitem{ando1} T. Ando and Y. Uemura: \jpsj {\bf 36} (1974) 959.
%
\bibitem{oksok} Y. Ono, H. Kuwano, K. Slevin, T. Ohtsuki and B. Kramer: 
J. Phys. Soc. Jpn. {\bf 62} (1993) 2762.  
%
\bibitem{liusarma} Dongzi Liu and S. Das Sarma: \prb {\bf 49} (1994) 2677.
\bibitem{onofuku} Y. Ono and S. Fukuda: \jpsj {\bf 61} (1992) 1676. 
%
\bibitem{mieck} B. Mieck: Europhysics Lett. {\bf 17} (1990) 453, Z. Phys. 
B{\bf 90} (1993) 427.
%
\bibitem{hucke94} B. Huckestein: \prl {\bf 72} (1994) 1080.
%
\bibitem{brody} T.A. Brody: Nuvo Cimento Lett. {\bf  7} (1973) 482.  
%
\bibitem{izrailev} F.M. Izrailev: Phys. Lett. {\bf A 134} (1988) 13; 
J. Phys. {\bf A 22} (1989) 865. 
\bibitem{kramack} B. Kramer and A. MacKinnon: in {\it Proc. Int. 
Seminar ^^ ^^ Localization in Disordered System''}, Johnsbach/b. Dresden, 1983 
(Teubner, Leibzig, 1984).
%
\bibitem{hucke92} B. Huckestein: Europhysics Lett. {\bf 20} (1992) 451.
%
\bibitem{argaman} N. Argaman, Y. Imry and U. Smilansky: \pr {\bf B 47}  
(1993) 4440.  
%
\bibitem{pandey} A. Pandey and M.L. Mehta: Commun. Math. Phys. {\bf 87}  
(1983) 449.
%
\bibitem{dupuis} N. Dupuis and G. Montambaux: \pr {\bf B43} (1991) 14390.
%
\bibitem{french} J.B. French, V.K. Kota, A. Pandey and S. Tomsovic: Ann. Phys. 
{\bf 181} (1988) 198 
%
\bibitem{altshklo} B.L. Al'tshuler and B.I. Shklovskii: Sov. Phys. JETP 
{\bf 64} (1986) 127.
%
\bibitem{foot1} Note that $N_{\rm mat}/10$ is the width of the energy window.
%
\bibitem{koch} S. Koch, R.J. Haug, K. v. Klitzing and K. Ploog: 
\prl {\bf 67} (1991) 883.
\bibitem{dolgopolov} V.T. Dolgopolov, G.V. Kravchenko, S.S. Shashkin and 
S.V. Kravchenko: \prb {\bf 46} (1992) 13303.
%
\bibitem{shashkin} A.A. Shashkin, V.T. Dolgopolov and G.V. Kravchenko: \prb 
{\bf 49} (1994) 14486.
%
\bibitem{shinohara} Y. Shinohara, T. Okamoto and S. Kawaji: unpublished.
%
\bibitem{schwzha} L. Schweitzer and I.Kh. Zharekeshev: J. Phys. Cond. Matt. 
{\bf 7} (1995) L377.
%
\bibitem{evange95} S.N. Evangelou: \prl {\bf 75} (1995) 2550.
\bibitem{hofstet2} E. Hofstetter and M. Schreiber: \prb {\bf 49} (1994) 14726.
\bibitem{zharek} I.Kh. Zharekeshev and B. Kramer: Jpn. J. Appl. Phys. (1995) 
in print.
%
\bibitem{foot2} In fact, a detailed observation of Fig.~\ref{Fig:7} tells us 
that the data for $x=0$ show a slight size dependence in the large $K$ region. 
This is thought to be due to the fact that the smaller the system size, the 
larger the original energy is if the value of $K$ is fixed; namely the effect 
of localization appears more strongly for smaller system sizes.  This 
consideration is consistent with the direction of the size dependence of the 
data for $x=0$ which is seen near $K \sim 30$ in Fig.~\ref{Fig:7}. 
\end{thebibliography}
\end{document}